
\input harvmac

\Title{UCLA/91/TEP/47 (revised)}{Composite Operators in
QCD$^*$}
\footnote{}{*This work was supported in part by the U.S.
Department
of Energy, under Contract DE-AT03-88ER 40384 Mod A006
Task C.}

\centerline{Hidenori SONODA}
\bigskip\centerline{\it Department of Physics,
UCLA, Los Angeles, CA 90024-1547, USA}

\vskip 1in
We give a formula for
the derivatives of a correlation function
of composite operators with respect to
the parameters (i.e., the strong fine structure
constant and the quark mass) of
QCD in four-dimensional euclidean space.
The formula is given as spatial integration of
the operator conjugate to a parameter.
The operator product of a composite
operator and a conjugate operator has
an unintegrable part, and the formula
requires divergent subtractions.
By imposing consistency conditions we derive
a relation between the anomalous
dimensions of the composite operators
and the unintegrable part of the operator
product coefficients.

\Date{May 1992}

\def\e{{\rm e}}
\def\exp{{\rm exp}}
\def\g1{g_{\bf 1}}
\def\one{{\bf 1}}
\def\Om{{\cal O}_m}
\def\OE{{\cal O}_E}
\def\dl{{d \over dl}}
\def\b1{{\beta_{\bf 1}}}
\def\bm{{\beta_m}}
\def\bE{{\beta_E}}
\def\vev#1{\langle #1 \rangle_{m,g_E}}
\def\O{{\cal O}}
\def\ep{\epsilon}
\def\gbarE{{\bar{g}_E}}
\def\mbar{{\bar{m}}}
\def\gbar1{{\bar{g}_\one}}
\def\On{\O_{i_1} (r_1) ... \O_{i_n} (r_n)}
\def\Onk{\O_{i_1} (r_1) ... \O_j (r_k) ... \O_{i_n} (r_n)}
\def\cbar{\bar{c}}

\def\t#1{\tilde{#1}}

\newsec{Introduction}

The dimensional regularization with
the minimal subtraction
has become everyone's favorite method
for perturbative calculations of renormalizable field theories.
\ref\rthooft{G. 't Hooft and M. Veltman, Nucl. Phys.
B44(1972)189\semi
C. G. Bollini and J. J. Giambiagi, Phys. Lett.
40B(1972)566\semi
J. F. Ashmore, Nuovo Cimento Lett. 4(1972)289}
The method has three advantages.
First, calculations are simple.  Secondly, it is mass
independent.
Thirdly, the singularities higher than $1/\ep$ are determined
completely by the simple $1/\ep$ poles.\ref\rthooft2{G. 't
Hooft, Nucl.
Phys. B62(1973)444}

Consider the $(\phi^4)_4$ theory as an example.  The theory

has three parameters $\g1$, $m^2$, and $\lambda$ with
scaling
dimensions $4$, $2$, and $0$, respectively.  These
parameters
are conjugate to the operators $\one$, $\phi^2/2$, and
$\phi^4/4!$ whose scaling dimensions are $0$, $2$, and
$4$, respectively.  Under the renormalization group (RG),
the coordinate distance $r$ transforms into $r \e^{- l}$,
while the renormalization point is always fixed at $r = 1$.
Hence, the RG acts toward the infrared limit, and it differs
from the standard definition by rescaling.   Then, under the
RG
the parameters transform as follows:
\eqn\ephi{\eqalign{\dl~\g1 &= 4 \g1 + {(m^2)^2 \over 2} \b1
(\lambda )\cr
\dl~ m^2 &= (2 + \bm (\lambda )) m^2\cr
\dl ~\lambda &= \beta_\lambda (\lambda ) .\cr}}
As we can see, the scaling dimensions of the parameters are
additively
preserved under the RG; only terms of scaling dimension
$4$ are allowed
in the first equation,
while only terms of scaling dimension $2, 0$ are allowed in
the second
and third, respectively.
In fact in ref.~\ref\rsonodarg{H.~Sonoda, Nucl. Phys.
B352[FS]
(1991)585} it was shown
that we can always choose parameters such that the scaling
dimensions of the parameters are additively preserved in
any field theory with an ultraviolet fixed point.  These
parameters
were also shown to be appropriate to describe the short
distance physics.
Locality of the theory implies that the RG equations involve
only integral powers of these parameters.  Hence, the RG
equations
become finite polynomials of dimensionful parameters
whose
coefficients are power series in dimensionless parameters.
Within perturbation theory the results of ref.~\rsonodarg\
applies also
to the $\phi^4$ theory.
The minimal subtraction with dimensional regularization
provides
an example of such a choice of parameters.
The structure of \ephi\ is called mass independent,
since the beta functions
depend only on the dimensionless parameter $\lambda$.

Now, in the minimal subtraction scheme in the dimensional
regularization,
the beta functions and the anomalous dimensions of
composite operators
are directly related to the simple $1/\ep$ poles of the
unrenormalized
correlation functions.  In other words, nontrivial
renormalization
properties of the theory, i.e., nonvanishing beta functions
and
anomalous dimensions, demand that
the bare correlation functions have simple $1/\ep$ poles.
Since the higher
order poles in $1/\ep$ are related to the simple $1/\ep$
poles, we can
say that all divergences in $\ep$ are {\it expected}
consequences of
nonvanishing beta functions and anomalous dimensions.

So far we have explained the advantages of the
dimensional regularization with the minimal subtraction.
We should not be totally happy
with this method, however, since the physical meaning of the
divergences in
$\ep$ is unclear.  The purpose of this paper is to consider
physical singularities of correlation functions at short
distances
and relate the short distance singularities, rather than
the unphysical singularities in $\ep$, to the anomalous
dimensions.
Our results are valid beyond perturbation theory, and
we will use QCD in four dimensional euclidean space as
an example.

The paper is organized as follows.  We summarize the
relevant facts on the RG equations in sect.~2 and
those on the operator product expansions
in sect.~3.  Then, we introduce the main formula in this
paper in sect.~4 that describes the change of correlation
functions of composite operators
under an infinitesimal change of parameters.
In sect.~5 we derive a relation
between the operator product coefficients and anomalous
dimensions of composite operators by considering
the consistency between the main formula
of sect.~4 and the RG eqs.  In sect.~6
we make yet another consistency check
of the main formula of sect.~4, coming from
commutativity of derivatives.  In sect.~7
we introduce a convention for composite operators.
We give concluding remarks in sect.~8.

\newsec{Renormalization group equations}

In this section we introduce relevant facts on the RG
equations in QCD with massive quarks in
four dimensional euclidean space.
The theory is characterized by three parameters $\g1$, $m$,
and $g_E$.  The parameter $\g1$ is an additive constant to
the
lagrangian
density, and its scaling dimension is four.
The parameter $m$ is the quark mass parameter
with scaling dimension one, and $g_E$ is
the strong fine structure constant with scaling dimension
zero.\foot{
We do not consider a gauge fixing parameter, since we will
only consider
gauge invariant operators in this paper.}
We denote the operators conjugate to these parameters
by $\one$, $\Om$, and $\OE$, respectively. The operator
$\one$ is the identity operator.
The operator $\Om$ is the mass density
operator $\bar{\psi} \psi$, and $\OE$ corresponds to
the energy density operator $F_{\mu\nu} F_{\mu\nu}$.
These three operators are the only gauge invariant scalar
operators with
scaling dimension less than or equal to four that conserve
C and P.  There is no other independent operator with these
properties; $\partial_\mu ( \bar{\psi} \gamma^\mu \psi ) = 0$
and $\bar{\psi}\gamma^\mu D_\mu \psi = m \bar{\psi} \psi$
by the eqs. of motion.

We can write down the RG eqs. in the following form:
\eqn\erg{\eqalign{\dl~\g1 &= 4 \g1 + {m^4 \over 4!} \b1 (g_E
)\cr
\dl~ m &= (1 + \bm (g_E ))m \cr
\dl~ g_E &= \bE (g_E ) .\cr }}
The above form is determined by the requirement that the
scaling
dimensions of the parameters are additively conserved
under
the RG.
In ref. \rsonodarg\ it was shown that we can always choose
such parameters.
In the first eq.  in \erg\ the coefficient of $\g1$ is simply 4,
since the
parameter $\g1$, being an additive constant in the
lagrangian,
has nothing to do with interactions.
Locality of the theory implies that the beta functions
can be expanded in powers of $g_E$ near $g_E = 0$:
\eqn\ebeta{\eqalign{\b1 (g_E ) &= \b1_{,0} + \b1_{,1} g_E + ...
\cr
\bm (g_E ) &= \bm_{,1} g_E + {\bm_{,2} \over 2} g_E^2 + ...
\cr
\bE (g_E ) &= {\bE_{,1} \over 2} g_E^2 + {\bE_{,2} \over 3!}
g_E^3 + ... ~.\cr}}

Let $F(\g1 , m, g_E )$ be the free energy density.  Then, by
definition
of the RG, we find
\eqn\ergf{\dl F = 4 F.}
Since
\eqn\evev{\vev{\one} = 1 = {\partial \over \partial \g1} F,~~~
\vev{\Om} = {\partial \over \partial m} F,~~~
\vev{\OE} = {\partial \over \partial g_E} F,}
we find, from \erg, \ergf, and \evev, that the composite
operators
$\Om$ and $\OE$ satisfy the RG eqs.
\eqn\ergop{\eqalign{\dl \Om &= (3 - \bm (g_E )) \Om
- {m^3 \over 3!} \b1 (g_E ) \one \cr
\dl \OE &= (4 - \bE '(g_E )) \OE - m \bm '(g_E ) \Om
- {m^4 \over 4!} \b1 '(g_E ) \one \cr}}
In principle there can be additional scalar operators whose
expectation
values vanish, for example total derivative operators.
But there is no other
scalar operators of dimension less than or equal to four,
and the above equations are exact.

We now introduce RG eqs. for general composite operators.

Let $\{ \O_i\}$
be a basis of gauge invariant scalar operators, where $\O_i$
has scaling
dimension $x_i$.  Then the RG eqs. take the form
\eqn\ergcomp{\dl \O_i = x_i \O_i + \sum_j \Gamma_{i,j}
(m,g_E ) \O_j,}
where the anomalous dimensions $\Gamma_{i,j}$ can be
nonvanishing only
if $x_i - x_j$ is a non-negative integer.  More precisely, the
anomalous
dimensions are finite polynomials of $m$ with coefficients as
power
series in $g_E$:
\eqn\egamma{\Gamma_{i,j} (m,g_E ) = \gamma_{i,j} (g_E )
m^{x_i - x_j},}
where $\gamma_{i,j} (g_E )$ is a power series.  We assume
that
\eqn\emix{\Gamma_{i,j} (0, 0) = 0}
so that at the UV fixed point $m=g_E=0$ there is no mixing
of
operators.  In the matrix notation,
in which $\O_i$ is an infinite dimensional column vector
$\O$, we can write
eqs. \ergcomp\ as
\eqn\ergmatrix{\dl \O = X \O + \Gamma (m,g_E ) \O,}
where $X_{i,j} = x_i \delta_{ij}$ is a diagonal matrix.

Before closing this section, we will examine the RG eqs.
\erg\ further
for later convenience.  We define the running parameters
$\gbarE (l;g_E ), \mbar (l;m,g_E )$, and $\gbar1 (l; \g1 , m,
g_E )$ as
the solutions of the RG eqs.
\eqn\egbar{\eqalign{{\partial \over \partial l} ~\gbarE (l;g_E )
&= \bE (\gbarE (l;g_E )) \cr
{\partial \over \partial l} ~\mbar (l;m,g_E )
&= (1 + \bm (\gbarE (l;g_E ))) \mbar (l;m,g_E ) \cr
{\partial \over \partial l} ~\gbar1 (l;\g1 , m, g_E )
&= 4 \gbar1 (l;\g1 , m, g_E )
+ {\mbar (l;m,g_E )^4 \over 4!} \b1 (\gbarE (l;g_E )),  \cr}}
with the initial conditions
\eqn\einitial{\gbarE (0;g_E ) = g_E ,~ \mbar (0;m,g_E ) = m,~
\gbar1 (0;\g1 , m, g_E ) = \g1 .}
We will suppress the initial conditions from now on, and we
will denote,
for example,
simply $\gbarE (l)$ instead of $\gbarE (l;g_E )$.
Then we note that
$$
\gbarE (\ln r),~ \mbar (\ln r),~ \gbar1 (\ln r)
$$
are all RG invariants, since the dependence on the
change of the coordinate $r$ under the RG
is canceled by the dependence on the change of
the initial parameters ${\g1},m, g_E$
under the RG.  Using the beta
functions we can write the running parameter $\mbar (l)$ as
follows:
\eqn\embar{\mbar (l) = m \e^l E(\gbarE (l), g_E ),}
where $E(y,x)$ is defined by
\eqn\eE{E(y, x) \equiv \exp \left[ \int_x^y dz ~
{\bm (z) \over \bE (z)} \right] .}
Note that the function $E$ satisfies
\eqn\eEeq{\bE (y) {\partial \over \partial y} E(y, x) = \bm (y)
E(y,x) ,~~~\bE (x) {\partial \over \partial x} E(y, x)
= - E(y,x) \bm (x) .}

\newsec{Operator product expansions}

In this section we summarize relevant facts on the operator
product
expansions (OPE).\ref\rwilson{K. Wilson, Phys. Rev.
179(1969)1499}
We will be interested in the two types of operator products:
\eqn\eope{\eqalign{\Om (r) \O (0) &= C_m (r;m,g_E ) \O (0)
+ o \left( {1 \over r^4} \right) \cr
\OE (r) \O (0) &= C_E (r;m,g_E ) \O (0) + o \left( {1 \over r^4}
\right) . \cr}}
Here, both $C_m$ and $C_E$ are part of the operator
coefficient functions that are at least as singular as $1/r^4$.
In taking the operator products
we have taken the average over the orientation of the
coordinate vector
$r_\mu$; without the average the right-hand sides of \eope\
would include
non-scalar composite operators.

One of the most fundamental properties of OPE is that for a
fixed coordinate
$r$ the coefficient functions can be expanded in powers of
the
small parameters $m, g_E$.  We will find that
this analyticity, together with the RG, implies that
the coefficient functions are finite polynomials
in $m$ whose coefficients are infinite power series in $g_E$,
i.e.,
\eqn\ecoeff{\eqalign{C_{m i,j} (r;m,g_E) &= \sum_{n = 0}^{x_i
- x_j -1}
{m^n \over n!} {\partial^n \over \partial m^n} C_{m i,j}
(r;0,g_E) \cr
C_{E i,j} (r;m,g_E) &= \sum_{n=0}^{x_i - x_j}
{m^n \over n!} {\partial^n \over \partial m^n} C_{E i,j} (r;0,g_E)
\cr }}
We note that this implies that $C_{m i,j}$ can be
nonvanishing
only for $x_i \ge x_j + 1$, while $C_{E i,j}$ can be
nonvanishing
only for $x_i \ge x_j$.  Let us derive the above results.

The coefficient functions are closed under the RG, and we
find
\eqn\ergcoeff{\eqalign{\dl ~C_m (r; m, g_E ) &=
(3 - \bm (g_E )) C_m (r; m, g_E )
+ [ X + \Gamma (m, g_E ), C_m (r; m, g_E )] \cr
\dl ~C_E (r; m, g_E )
&= (4 - \bE' (g_E )) C_E (r; m, g_E ) - m \bm' (g_E ) C_m
(r;m, g_E ) \cr
&~~~~~ + [ X + \Gamma (m, g_E ), C_E (r; m, g_E )] . \cr}}
In order to solve these RG eqs. we introduce a matrix
$G(r;m,g_E )$ that
satisfies
\eqn\eGeq{\dl G(r;m,g_E ) = \left( X + \Gamma (m,g_E )
\right) G(r;m,g_E )}
and the initial condition $G(1;m,g_E ) = 1$.  The solution is
given by
\eqn\eG{G(r;m,g_E ) =
{1 \over r^X} ~{\cal T} \exp \left[
\int_{\gbarE (\ln r)}^{g_E} dx~
{1 \over \bE (x)} \Gamma (\mbar (\ln r) E(x,\gbarE (\ln r)), x)
\right] ,}
where ${\cal T}$ denotes the increasing ordering of $x$ from
right to left.
Due to eq. \egamma\ $\Gamma_{i,j} (m,g_E) =
m^{x_i - x_j} \gamma_{i,j} (g_E)$, the matrix element $G_{i,j}
(r;m,g_E)$
can be nonvanishing only if $x_i - x_j$ is a non-negative
integer.
As far as the power of $r$ is concerned (i.e., we ignore
logarithmic corrections),
$\mbar (\ln r )$ is proportional to $r$ from eq. \embar.
Hence the dependence of $G_{i,j} (r;m,g_E)$ on the powers
of
$m$ and $r$ is given by
\eqn\eGorder{G_{i,j} (r;m,g_E) \propto {m^{x_i - x_j} \over
r^{x_j}}.}
Similarly, we find
\eqn\eGordertwo{G^{-1}_{i,j} (r;m,g_E) \propto m^{x_i - x_j}
r^{x_i}.}
It is also helpful to note
that the matrix $G(r;m,g_E )$ satisfies
\eqn\eGeqtwo{{\partial \over \partial \ln r} G(r;m,g_E ) = -
G(r;m,g_E )
\big( X + \Gamma (\mbar (\ln r), \gbarE (\ln r) ) \big) .}

Using the matrix $G$ we can solve the RG eqs. \ergcoeff\ as

follows:
\eqn\ecoeffsol{\eqalign{
C_m (r;m,g_E) &= {1 \over r^3} E(\gbarE (\ln r), g_E)
G(r;m,g_E)
H_m (\mbar (\ln r), \gbarE (\ln r)) G^{-1} (r;m,g_E) \cr
C_E (r;m,g_E) &= {1 \over r^4} {\bE (\gbarE (\ln r)) \over \bE
(g_E)}
G(r;m,g_E) H_E (\mbar (\ln r), \gbarE (\ln r)) G^{-1} (r;m,g_E)
\cr
&~~~+ {m \over \bE (g_E)} (\bm (\gbarE (\ln r))- \bm (g_E))
C_m (r;m,g_E) ,\cr}}
where we define
\eqn\eH{H_m (m,g_E) \equiv C_m (1;m,g_E),~~~
H_E (m,g_E) \equiv C_E (1;m,g_E).}
{}From the analyticity assumption, both
$H_m (m,g_E)$ and $H_E (m,g_E)$ are
power series in $m$ and $g_E$.  Suppose $H_{m i,j}
(m,g_E)$
has a term proportional to $m^n$.  Then, from eq. \ecoeffsol\

we find that $C_{m i,j} (r;m,g_E)$ has a term proportional to
$m^n/r^{3+x_i - x_j -n}$.  Since $C_m (r;m,g_E)$ must be
at least as singular as $1/r^4$ by definition, we must have
$n \le x_i - x_j -1$.  Hence, $H_{m i,j} (m,g_E)$ is
a polynomial of degree $x_i - x_j -1$ as far as $m$ is
concerned.
The $m^n$ term in $H_{m i,j}$ can also contribute to
$C_{m i',j'} (r;m,g_E)$, where $x_i' \ge x_i > x_j \ge x_j'$.
But from \eGorder\ and \eGordertwo\ we find again
that $n \le x_i - x_j -1$.
Similarly, we can conclude that $H_{E i,j} (m,g_E)$ is
a polynomial of degree $x_i - x_j$ with respect to $m$.
To summarize, we find
the structure
\eqn\eHpoly{\eqalign{H_{m i,j} (m,g_E) &= \sum_{n = 0}^{x_i -
x_j -1}
{m^n \over n!} H_{m i,j}^{(n)} (g_E) ,\cr
H_{E i,j} (m,g_E) &= \sum_{n = 0}^{x_i - x_j}
{m^n \over n!} H_{E i,j}^{(n)} (g_E) .\cr}}

\newsec{Variational formulas}

We will introduce a main formula that gives operator
realization of the
derivatives with respect to $m$ and $g_E$ in this section.
We will
see that the consistency of this formula with the RG eqs.
gives a relation between the operator coefficients and
anomalous dimensions.

So far we have been saying casually that the parameter $m$
and
the operator $\Om$ are {\it conjugate} to each other.  (The
following discussion applies to $g_E$ and $\OE$ as well.)
The precise meaning of the conjugacy relation is
that the derivative $- {\partial \over \partial m}$ is realized
by the operator $\Om$.  One example is eq. \evev.
This conjugacy relation must be valid not only for
the simple expectation value as \evev\ but also for any
correlation functions of composite operators.  So, we
expect
a formula of the type
\eqn\evarpre{ - {\partial \over \partial m} \vev{\On} =
\int d^4 r~ \vev{\Om (r) \On}^c ,}
where
\eqn\econn{\vev{\Om (r) \On}^c \equiv \vev{ (\Om (r) -
\vev{\Om})
\On} .}
We immediately notice, however, that the above formula is
not
well defined due to the short distance singularities in the
product of $\Om$ and $\O_i$.  We emphasize that these
short distance singularities
are physical singularities that exist in QCD
even after renormalization.  (We will make a brief
remark on the relation between these short distance
singularities and the singularities we encounter
in perturbative calculations in the concluding section.)
There must be a way of fixing
this problem in a local way, since the singularities result
from the UV physics rather than the IR physics.  Thus, we
postulate
the validity of the following variational formulas in QCD
(an analogous formula was mentioned in ref. \rwilson):
\eqn\evar{\eqalign{
& - {\partial \over \partial m} \vev{\On} =
\vev{\Om^* \On} \cr
& ~~~~~~~~ \equiv
\lim_{\ep \to 0}
\Big[ \int_{|r - r_i| \ge \ep} d^4 r~ \vev{\Om (r) \On}^c \cr
&~~~~~~~~~~ + \sum_{k=1}^n
A_{m i_k,j} (\ep ; m,g_E) \vev{\Onk} \Big] \cr
& - {\partial \over \partial g_E} \vev{\On} =
\vev{\OE^* \On} \cr
& ~~~~~~~~ \equiv \lim_{\ep \to 0}
\Big[ \int_{|r - r_i| \ge \ep} d^4 r~ \vev{\OE (r) \On}^c \cr
&~~~~~~~~~~ + \sum_{k=1}^n
A_{E i_k,j} (\ep ; m,g_E) \vev{\Onk} \Big] , \cr}}
where we define
\eqn\eA{\eqalign{A_{m i,j} (\ep ; m,g_E) &\equiv
- \int_{1 \ge r \ge \ep} d^4 r~ C_{m i,j} (r;m,g_E) + c_{m i,j}
(m,g_E) \cr
A_{E i,j} (\ep ; m,g_E) &\equiv
- \int_{1 \ge r \ge \ep} d^4 r~ C_{E i,j} (r;m,g_E) + c_{E i,j}
(m,g_E) .\cr}}
The variational formulas \evar\ explain why it is important
to treat the operator coefficients at least as singular as
$1/r^4$;
less singular coefficients are irrelevant for the subtractions.
The finite counterterms $c_m$, $c_E$ are necessary
in order to compensate the arbitrariness involved in the
subtracting procedure.  The finite

counterterms must have the same structure
as the operator coefficients $C_m$, $C_E$ (for a fixed $r$),
and
they can be expanded as
\eqn\ec{\eqalign{c_{m i,j} (m,g_E) &= \sum_{n=0}^{x_i - x_j
-1}
{m^n \over n!} c_{m i,j}^{(n)} (g_E) \cr
c_{E i,j} (m,g_E) &= \sum_{n=0}^{x_i - x_j}
{m^n \over n!} c_{E i,j}^{(n)} (g_E) .\cr}}
For later convenience we define the maximal parts of the
counterterms by
\eqn\ecmax{\eqalign{\cbar_{m i,j} (m,g_E) &\equiv {m^{x_i -
x_j -1}
\over (x_i - x_j -1)!} ~c_{m i,j}^{(x_i - x_j -1)} (g_E) \cr
\cbar_{E i,j} (m,g_E) &\equiv {m^{x_i - x_j}
\over (x_i - x_j)!} ~c_{E i,j}^{(x_i - x_j)} (g_E) .\cr}}

\newsec{Consistency condition}

It is very important to note that the variational formulas
introduced
in the previous section are assumed to be valid for finite
parameters
$m, g_E$, i.e., they are assumed to be valid
beyond perturbation theory.  Though it is difficult to derive
the variational formulas from first principles,
it is easy to check their consistency.  Especially
we will check their consistency with the RG eqs.

We first examine
the transformation of the left-hand sides of the variational
formulas
\evar\ under the RG.  From eqs. \erg, \ergop, and \ergmatrix\
we find
\eqn\eleftm{\eqalign{& \dl \left( - {\partial \over \partial m}
\vev{\On} \right) \cr
&~~ = \sum_{k=1}^n \left(X_{i_k,j} + \Gamma_{i_k,j} (m,g_E)
\right)
\left( - {\partial \over \partial m} \vev{\Onk} \right) \cr
&~~~ - (1+\bm (g_E))
\left( - {\partial \over \partial m} \vev{\On} \right) \cr
&~~~ - \sum_{k=1}^n \partial_m \Gamma_{i_k,j} (m,g_E)
\vev{\Onk} \cr}}
and
\eqn\eleftE{\eqalign{& \dl \left( - {\partial \over \partial g_E}
\vev{\On} \right) \cr
&~~~ = \sum_{k=1}^n \left( X_{i_k,j} + \Gamma_{i_k,j}
(m,g_E) \right)
\left( - {\partial \over \partial g_E} \vev{\Onk} \right) \cr
&~~~ - \bE' (g_E) \left( - {\partial \over \partial g_E} \vev{\On}
\right) \cr
&~~~ - m \bm' (g_E) \left( - {\partial \over \partial m} \vev{\On}
\right) \cr
&~~~ - \sum_{k=1}^n \partial_{g_E} \Gamma_{i_k,j} (m,g_E)
\vev{\Onk} .\cr}}

Next we consider the transformation of the right-hand sides
of \evar.
{}From eqs. \ergop\ and \ergmatrix\ we obtain
\eqn\erightm{\eqalign{& \dl \vev{\Om^* \On} \cr
&~~ = \sum_{k=1}^n
\left( X_{i_k,j} + \Gamma_{i_k,j} (m,g_E) \right) \vev{\Om^*
\Onk} \cr
&~~~ - (1+\bm (g_E)) \vev{\Om^* \On} \cr
&~~~ - 2 \pi^2 \sum_{k=1}^n H_{m i_k,j} (m,g_E) \vev{\Onk}
\cr
&~~~ + \sum_{k=1}^n \left( \dl c_{m i_k,j}
+ (1+\bm (g_E)) c_{m i_k,j} +
\left[ c_m , X + \Gamma (m,g_E) \right]_{i_k,j} \right) \cr
&~~~~~~~~~~ \cdot \vev{\Onk} \cr}}
and
\eqn\erightE{\eqalign{& \dl \vev{\OE^* \On} \cr
&~~ = \sum_{k=1}^n
\left( X_{i_k,j} + \Gamma_{i_k,j} (m,g_E) \right) \vev{\OE^*
\Onk} \cr
&~~~ - \bE' (g_E) \vev{\OE^* \On} - m \bm' (g_E) \vev{\Om^*
\On} \cr
&~~~ - 2 \pi^2 \sum_{k=1}^n H_{E i_k,j} (m,g_E) \vev{\Onk}
\cr
&~~~ + \sum_{k=1}^n \left( \dl c_{E i_k,j} + \bE' (g_E) c_{E
i_k,j}
+ m \bm' (g_E) c_{m i_k,j} +
\left[ c_E, X + \Gamma (m,g_E) \right]_{i_k,j} \right) \cr
&~~~~~~~~~ \cdot \vev{\Onk} . \cr}}
Here we get the $H_m, H_E$ terms from the boundary
integrals at
$r = 1$, and $2 \pi^2$ is the volume integral of a unit
3-sphere.

By comparing eq. \eleftm\ and eq. \erightm, and eq. \eleftE\
and eq. \erightE\ we obtain
\eqn\econsist{\eqalign{2 \pi^2 H_m (m,g_E) &= \partial_m
\Gamma (m,g_E)
+ \dl c_m (m,g_E) \cr
&~~~ + (1 + \bm (g_E)) c_m (m,g_E)
+ [ c_m (m,g_E), X + \Gamma (m,g_E) ] \cr
2 \pi^2 H_E (m,g_E) &= \partial_{g_E} \Gamma (m,g_E)
+ \dl c_E (m,g_E) \cr
&~~~ + \bE' (g_E) c_E (m,g_E) + m \bm' (g_E) c_m (m,g_E)
\cr
&~~~ + [ c_E (m,g_E), X + \Gamma (m,g_E) ] . \cr}}
These formulas give relations between the operator
coefficients
and anomalous dimensions.   We can rewrite these in a more
suggestive way as follows:
\eqn\econsisttwo{\eqalign{2 \pi^2 H_m &=
\partial_m \Phi + [c_m, \Phi ] + \bE (\partial_{g_E} c_m -
\partial_m c_E + [c_E, c_m]) \cr
2 \pi^2 H_E &= \partial_{g_E} \Phi + [c_E, \Phi ]
- m (1+\bm ) (\partial_{g_E} c_m -
\partial_m c_E + [c_E, c_m]) ,\cr}}
where
\eqn\ePhi{\Phi (m,g_E) \equiv X + \Gamma + \bE c_E +
m (1 + \bm ) c_m .}
Eqs.~\econsisttwo\ constitute a main result of this paper,
and they are analogous to the relation between the
anomalous
dimensions and $1/\ep$ poles in the dimensional
regularization
with the minimal subtraction.

We will come back to the geometric meaning of the
expressions
\econsisttwo\ in sect.~7.  Note that the anomalous

dimensions $\Gamma$ contribute only to the maximal
parts of the coefficient functions.

Now, by substituting eqs. \econsist\ into eqs. \ecoeffsol\
we can express the coefficient functions
for an arbitrary $r$ as total derivatives:
\eqn\ecoeffder{\eqalign{& C_m (r;m,g_E) \cr
&~~ =
{1 \over 2 \pi^2 m r^4} {\partial \over \partial \ln r} \Bigg(
G(r;m,g_E) \Big( X + \mbar (\ln r) c_m (\mbar (\ln r), \gbarE
(\ln r)) \Big)
G^{-1} (r;m,g_E) \Bigg) \cr
& C_E (r;m,g_E) \cr
&~~ =
{1 \over 2 \pi^2  \bE (g_E) r^4} {\partial \over \partial \ln r}
\Bigg(
G(r;m,g_E) \Big( \Gamma (\mbar (\ln r), \gbarE (\ln r))
- \bm (g_E) X \cr
&~~~~~~~~~~ + \bE (g_E(\ln r)) c_E (\mbar (\ln r), \gbarE (\ln
r)) \cr
&~~~~~ + \mbar (\ln r) \big(\bm (\gbarE (\ln r)) - \bm
(g_E)\big)
c_m (\mbar (\ln r), \gbarE (\ln r)) \Big) G^{-1} (r;m,g_E) \Bigg)
. \cr}}
To derive this result we need to use
eqs. \embar, \eGeqtwo, and
\eqn\exgamma{[X, \Gamma (m,g_E)] = m \partial_m \Gamma
(m,g_E),}
where the last equation is a consequence of eq. \egamma.

We can now integrate the above coefficient functions
to obtain an expression for the subtractions
$A_m$ and $A_E$, defined by \eA, in terms of the
anomalous dimensions
$\Gamma$ and the counterterms $c_m, c_E$.
We find
\eqn\eAfinal{\eqalign{& A_m (\ep ;m,g_E) \cr
&~~ = {1 \over m}
\Big(  - X + G(\ep ;m,g_E) X G^{-1} (\ep ;m,g_E)  \cr
&~~~~~ +
G(\ep ;m,g_E) \mbar (\ln \ep ) c_m (\mbar (\ln \ep ), \gbarE
(\ln \ep ))
G^{-1} (\ep ;m,g_E) \Big), \cr
& A_E (\ep ;m,g_E) \cr
&~~ = {1 \over \bE (g_E)} \Bigg(
\bm (g_E) (X - G X G^{-1}) - \Gamma (m,g_E) +
G \Gamma (\mbar (\ln \ep ), \gbarE (\ln \ep )) G^{-1} \cr
&~~~~~ + G \bigg( \bE (\gbarE (\ln \ep ))
c_E (\mbar (\ln \ep ), \gbarE (\ln \ep )) \cr
&~~~~~~~~~~ + \big(\bm (\gbarE (\ln \ep )) - \bm (g_E)\big)
\mbar (\ln \ep ) c_m (\mbar (\ln \ep ), g_E(\ln \ep ))
\bigg) G^{-1} \Bigg). \cr}}

Using the above formulas we can count the powers of
$1/\ep$ in the
subtractions.  First we note, from \eGorder\ and
\eGordertwo, that
\eqn\eXGorder{\eqalign{(X - G X G^{-1})_{i,j} &= 0,
~~~~~
{\rm if}~x_i = x_j \cr
&\propto m^{x_i - x_j},~~~{\rm if}~x_i > x_j .\cr}}
Since $G(1;m,g_E) = 1$, the term $(X - G X G^{-1})_{i,j}$
corresponds
to purely logarithmic contributions in $\ep$ that vanish at $\ln
\ep =0$.
Similarly, we find that
\eqn\eGGamma{\Big(\Gamma (m,g_E) - G \Gamma (\mbar
(\ln \ep),
\gbarE (\ln \ep)) G^{-1}\Big)_{i,j} \propto m^{x_i - x_j}}
are purely logarithmic in $\ep$ without powers of $\ep$.
Secondly we note, from \ec, that
\eqn\eGc{\eqalign{\Big(G \mbar (\ln \ep) c_m (\mbar (\ln \ep),
\gbarE (\ln \ep )) G^{-1}\Big)_{i,j}
&\sim c_{m i',j'}^{(n)} (\gbarE (\ln \ep ))
 { (m \ep )^{n+1} \over \ep^{x_i - x_j}} \cr
\Big(G c_E (\mbar (\ln \ep), \gbarE (\ln \ep )) G^{-1}\Big)_{i,j}
&\sim c_{E i',j'}^{(n)} (\gbarE (\ln \ep ))
{ (m \ep )^n \over \ep^{x_i - x_j}} , \cr}}
where $x_i \ge x_{i'} \ge x_{j'} \ge x_j$.  Especially we find
\eqn\eGcbar{\eqalign{\Big(G \mbar (\ln \ep)
\cbar_m (\mbar (\ln \ep), \gbarE (\ln \ep ))
G^{-1}\Big)_{i,j}
&\sim \cbar_{m i',j'} (\gbarE (\ln \ep ))
m^{x_i - x_j} \cr
\Big(G \cbar_E (\mbar (\ln \ep), \gbarE (\ln \ep ))
G^{-1}\Big)_{i,j}
&\sim \cbar_{E i',j'} (\gbarE (\ln \ep ))
m^{x_i - x_j} . \cr}}
These are not necessarily zero at $\ep =1$ and involve terms

that are finite as $\ep \to 0$.  In conclusion the
anomalous dimensions
$\Gamma$ give only logarithmic contributions to the
subtractions
$A_m, A_E$, while the non-maximal parts of the
counterterms $c_m, c_E$
give contributions that are proportional to integral powers of
$1/\ep$,
and the maximal parts
$\cbar_m, \cbar_E$ give both logarithmic and finite
contributions
in $\ep$.  In the absence of maximal counterterms there is no

finite subtraction.  Thus, the necessary subtractions are
minimal when
the maximal parts of the counterterms vanish.

\newsec{Commutativity}

In the previous section we checked the consistency of the
variational
formulas \evar\ with the RG equations.  In this section we will
perform
another consistency check.  The variational formulas give
single
derivatives of correlation functions with respect to $m, g_E$,

and we can use the formulas recursively to realize
multiple derivatives in terms of multiple integrals over
operators.

We define two insertions $\vev{\OE^* \Om^* ...}$ (in this
particular
order) such that it equals
$- \partial_{g_E} \vev{\Om^* ...}$; namely
we apply the definition
of $\OE^*$ to the correlation function in the definition of
$\vev{\Om^* ...}$:
\eqn\eEm{\eqalign{& \vev{\OE^* \Om^* \On} \equiv
\lim_{\ep \to 0} \Bigg[ \int_{|r - r_i| \ge \ep} d^4 r ~
\vev{\OE^* \Om (r) \On}^c \cr
& ~~~~~ + \sum_{k=1}^n A_{m i_k, j} (\ep ;m,g_E)
\vev{\OE^* \Onk} \cr
& ~~~~~ - \sum_{k=1}^n \partial_{g_E} A_{m i_k,j} (\ep
;m,g_E)
\vev{\Onk} \Bigg] .\cr}}
Similarly, we define $\vev{\Om^* \OE^* \O ... }$ such that
it equals $- \partial_m \vev{\OE^* \O ... }$:
\eqn\emE{\eqalign{& \vev{\Om^* \OE^* \On} \equiv
\lim_{\ep \to 0} \Bigg[ \int_{|r' - r_i| \ge \ep} d^4 r' ~
\vev{\Om^* \OE (r') \On}^c \cr
& ~~~~~ + \sum_{k=1}^n A_{E i_k, j} (\ep ;m,g_E)
\vev{\Om^* \Onk} \cr
& ~~~~~ - \sum_{k=1}^n \partial_m A_{E i_k,j} (\ep ;m,g_E)
\vev{\Onk} \Bigg] .\cr}}
In a similar way we can define $\vev{\Om^* \Om^* ...}$
and $\vev{\OE^* \OE^* ... }$.

The insertions $\Om^*$ and $\OE^*$ are operator
realizations
of the partial derivatives $- \partial_m$ and $-
\partial_{g_E}$.
Since the partial derivatives commute with one
another, we must satisfy
\eqn\ecomm{\vev{\OE^* \Om^* \On} = \vev{\Om^* \OE^*
\On}.}
The evaluation of the difference $\OE^* \Om^* - \Om^*
\OE^*$ is
straightforward but tedious,

and we will sketch the calculation in Appendix A.
For simplicity we give the result only for the case $n = 1$
(For the case of a general $n$, see Appendix A.):
\eqn\ecommfinal{\eqalign{& \vev{(\OE^* \Om^* - \Om^*
\OE^*) \O_i (0)} \cr
&~~~~~ = (\Omega_{Em})_{i,j} \vev{\O_j} -
(\partial_{g_E} c_m - \partial_m c_E + [ c_E, c_m ])_{i,j}
\vev{\O_j} .\cr}}
Here we define the curvature
\eqn\eOmega{\eqalign{& (\Omega_{Em})_{i,j} \vev{\O_j} \cr
& ~~~ \equiv \int_{1 \ge r} d^4 r~ {\rm F.P.} \int_{1 \ge r'} d^4
r'~
\langle \Om (r) (\OE (r') \O_i (0) - C_{E i,j} (r';m,g_E) \O_j (0))
\cr
& ~~~~~~~~~~~~~~ - \OE (r) (\Om (r') \O_i (0) - C_{m i,j}
(r';m,g_E) \O_j (0))
\rangle_{m,g_E} ,\cr}}
where {F.P.} denotes an integrable part with respect to $r$.

Thus, the consistency condition becomes
\eqn\ecurv{\partial_{g_E} c_m - \partial_m c_E + [ c_E, c_m ]
= \Omega_{Em}.}
The left-hand side is reminiscent
of a gauge field strength with $c_m, c_E$ as the gauge
fields.
We will elaborate on this view-point in the next section.
It is important to emphasize that

the curvature $\Omega_{Em}$ is related
essentially to a product of three operators close together,
and it cannot be deduced from the OPE coefficients which
have
to do with the products of only two operators close
together.

Before we close this section we must ask if

we get further consistency conditions from more

number of insertions of $\Om^*$ and $\OE^*$.
Let us consider insertion of K $\Om^*$'s and L
$\OE^*$'s.  These are defined using \evar\ recursively.
We can classify all possible insertions into two classes,
those of type 1
$$
\vev{\Om^* \O^* ... \On}
$$
and those of type 2
$$
\vev{\OE^* \O^* ... \On} ,
$$
where $\O^* ...$ corresponds to the remaining $K+L -1$
insertions.
Suppose that the insertion of $K'$ $\Om^*$'s and
$L'$ $\OE^*$'s is independent of ordering if
$K' + L' < K + L$.  Now, using the variational formulas \evar\
{\it only once},
we get
\eqn\emultider{\eqalign{\vev{\Om^* \O^* ... \On} &=
- \partial_m \vev{\O^* ... \On} \cr
\vev{\OE^* \O^* ... \On} &= - \partial_{g_E} \vev{\O^* ... \On}
.\cr}}
Hence, by the assumption all correlation functions are equal
in each class.  The equality between the two classes
amounts
to show
\eqn\eamount{\vev{\Om^* \OE^* \O^* ... \On}
= \vev{\OE^* \Om^* \O^* ... \On}}
where $\O^* ...$ stands for the remaining $K+L-2$ insertions.
Since $\vev{\O^* ... \On}$ is given as an integral
over $K+L-2+n$-point correlation function (ignoring
counterterms for simplicity),
we can prove the equality by going
through the same proof as given in Appendix A for
the case of two insertions.  Therefore, we will not get any
further
consistency condition from multiple insertions.

\newsec{Unique convention for composite operators}

We would like to redefine the composite operators in such a
way
that the corresponding counterterms have vanishing
maximal parts.
The reason is that this convention is the simplest in which
finite subtractions
are unnecessary.

The most general redefinition of the operators $\{ \O_i \}$
that we can
think of is given by
\eqn\eredef{\tilde{\O_i} = \sum_{j} N_{i,j} (m,g_E) \O_j,}
where the matrix $N$ is invertible.  The element $N_{i,j}$ can

be nonvanishing only if $x_i - x_j$ is a non-negative integer,
and $N_{i,j}$ must be proportional to $m^{x_i - x_j}$:
\eqn\edN{\partial_m N = {1 \over m} [X,N] .}
This condition is required in order to preserve the scaling
dimensions
of the redefined operators additively under the RG.  With
respect to $g_E$,
the element $N_{i,j}$ must be a power series as required
by locality of the theory.

For the redefined operators $\tilde{\O}$, the counterterms for

the variational formulas \evar\ are given by
\eqn\ectilde{\eqalign{\t{c}_m &= N c_m N^{-1}
- \partial_m N \cdot N^{-1} \cr
\t{c}_E &= N c_E N^{-1} - \partial_{g_E} N \cdot N^{-1} .}}
Hence, the finite counterterms transform as gauge fields
in the space of parameters $m, g_E$.  The inhomogeneous
terms are maximal, and for the maximal parts
alone we find
\eqn\ectildemax{\eqalign{\bar{\t{c}}_m &= N \cbar_m N^{-1}
- \partial_m N \cdot N^{-1} \cr
\bar{\t{c}}_E &= N \cbar_E N^{-1} - \partial_{g_E} N \cdot
N^{-1} .}}

The transformation properties \ectilde\ imply that we can
interpret $(c_m, c_E)$ as a gauge field on the
two-dimensional
theory space whose coordinates are $m$ and $g_E$.
The operators $\{\O_i\}$ form a basis of an infinite
dimensional
vector bundle.  The commutativity condition \ecurv\

gives the curvature of the gauge field in terms of an
integral over a correlation function \eOmega.  Since the
integral
has no reason to vanish, the curvature is nonvanishing
in general.  The field $\Phi$ defined by \ePhi\ can be

seen to transform as
\eqn\ePhitrans{\tilde{\Phi} = N \Phi N^{-1} .}
Thus, $\Phi$ is a scalar field in the adjoint representation
on the theory space.  Eqs.~\econsisttwo\ and \ecurv\
give the vector
field $(H_m, H_E)$ in terms of a covariant derivative
of $\Phi$ and the curvature $\Omega_{Em}$.

Hence, $(H_m, H_E)$ is
in the adjoint representation as expected.
So much for geometric interpretations of our results.

The simplest convention for the finite counterterms is the
vanishing maximal counterterms:
\eqn\ecbarzero{\bar{\t{c}}_m = 0,~~~ \bar{\t{c}}_E = 0.}
But in general this is impossible, since the curvature
$\Omega_{Em}$, defined by \eOmega, is nonvanishing.
We remove the ambiguity of the convention by imposing
\eqn\ecEzero{\bar{\t{c}}_E (m,g_E) = 0}
and
\eqn\ecmzero{\bar{\t{c}}_m (m, g_E = 0) = 0 .}
This is the analogue of the temporal gauge used in
non-abelian gauge theories.
For completeness we will derive the matrix of transformation
$N$ in Appendix B.  The gauge conditions \ecEzero\
and \ecmzero\ fix $N$ up to a left multiplication by
a constant matrix.  Hence,
the choice of the composite
operators is uniquely determined up to a

left multiplication by a constant matrix.

Finally we ask whether the unique convention we obtained
depends
on the choice of the renormalized parameters $m, g_E$.
The most general redefinition of the parameters are given by

\eqn\erepara{\eqalign{\t{m} &= m f(g_E) \cr
\t{g_E} &= g(g_E) ,\cr}}
where
\eqn\eefg{\eqalign{f(g_E) &= 1 + f_1 g_E + {f_2 \over 2}
g_E^2 + ... \cr
g(g_E) &= g_E + {g_1 \over 2} g_E^2 + {g_2 \over 3!} g_E^3
+ ...~.\cr}}
The corresponding changes in the partial derivatives are
\eqn\edertrans{\eqalign{\partial_{\t{m}} &= {1 \over f(g_E)}
{}~\partial_m \cr
\partial_{\t{g_E}} &= {1 \over g'(g_E)} ~\partial_{g_E} - m
{f'(g_E) \over
g'(g_E) f(g_E)} ~\partial_m .}}
The operators $\O_m$ and $\O_E$ transform in the same
way as
a vector:
\eqn\eOtrans{\eqalign{\t{\O}_m &= {1 \over f(g_E)} ~\Om \cr
\t{\O}_E &= {1 \over g'(g_E)} ~\OE - m {f'(g_E) \over
g'(g_E) f(g_E)} ~\Om .}}
This induces homogeneous transformations of the operator
coefficients:
\eqn\eCtrans{\eqalign{\t{C}_m (r;\t{m},\t{g_E})
&= {1 \over f(g_E)} ~C_m (r;m,g_E) \cr
\t{C}_E (r;\t{m},\t{g_E}) &= {1 \over g'(g_E)} ~C_E (r;m,g_E)
- m {f' (g_E) \over g'(g_E) f(g_E)} ~C_m (r;m,g_E) .}}
{}From eqs. \edertrans\ and \eCtrans\ we find that the
counterterms
also transform homogeneously as a vector:
\eqn\ectrans{\eqalign{\t{c}_m (\t{m},\t{g_E})
&= {1 \over f(g_E)} ~c_m (m,g_E) \cr
\t{c}_E (\t{m},\t{g_E}) &= {1 \over g'(g_E)}~ c_E (m,g_E)
- m {f' (g_E) \over g'(g_E) f(g_E)} ~c_m (m,g_E) .}}
Therefore, under the change of parameters the gauge
condition
\ecEzero\ is {\it not} preserved, while the condition \ecmzero\
is preserved.
Thus, in order to remove the ambiguity in the choice of
composite
operators by the gauge conditions \ecEzero\ and \ecmzero,
we
need to introduce a convention for the choice of $m$ and
$g_E$.
This can be done by applying conditions analogous to
\ecEzero,
\ecmzero\ to the conjugate operators $\Om, \OE$.

\newsec{Concluding remarks}

In this paper we obtained two main results.  First we
obtained
the variational formulas \evar\ that realize the partial
derivatives
with respect to $m, g_E$ in terms of the operator
insertions $\Om^*, \OE^*$.  Secondly we found a
relation between the anomalous dimensions of
composite operators and
their operator product coefficients, given by
eqs.~\econsisttwo.

Using the second result we can compute the subtractions
$A_m, A_E$, which are necessary in the variational formulas
\evar, in terms of the beta functions, anomalous dimensions
$\Gamma$, and finite counterterms $c_m, c_E$.

The finite counterterms are not unique, since
the composite operators are susceptible to
redefinitions \eredef.
Once we remove this ambiguity by imposing
conditions \ecEzero\ and \ecmzero,
the finite counterterms become unique, and
they are determined by eqs.~\econsisttwo\ and

the commutativity condition \ecurv.

The non-maximal part of the
finite counterterms corresponds to
subtractions divergent like powers of the cutoff
distance $\ep$.  (See the last paragraph
of sect.~5.)  The dimensional regularization
only sees logarithmic divergences and has no
counterpart to the non-maximal finite counterterms.
The maximal part of the finite counterterms gives
finite subtractions in the variational formulas.
This results from our asymmetric treatment of
conjugate operators when we evaluate
higher order derivatives of correlation functions
with respect to parameters.  In the dimensional
regularization each operator conjugate to a parameter
is treated symmetrically.

We would like to emphasize the conceptual
aspects of our results rather than their potential
usefulness
for perturbative calculations.  In fact some
modifications
will be necessary to apply
the variational formulas \evar\ for perturbative
calculations of the correlation functions of composite
operators.  The reason is that the operator
$\OE$, conjugate to $g_E$, roughly corresponds to
\eqn\eOEF{\OE \sim
{1 \over g_E^2} ~\rm{tr}~ (F_{\mu \nu})^2 ,}
where the field strength is defined by
\eqn\eFA{F_{\mu \nu} \equiv \partial_\mu A_\nu - \partial_\nu
A_\mu
+ [A_\mu , A_\nu ]}
in terms of  the gauge field, since
the lagrangian is given by
\eqn\elag{L = {1 \over g_E} ~\rm{tr}~(F_{\mu \nu})^2 .}
Note that this is not an artifact of normalization.  Even if we
redefine
a new $A$ by $\sqrt{g_E} A$, we still get $\OE \sim
F^2/g_E$.
The form of the operator $\OE$ given by \eOEF\ implies
that the correlation functions involving $\OE$ in fact
diverge in the limit $g_E \to 0$.  (But this does not
upset our assumption of the analyticity of the
operator coefficients $C_m, C_E$ with respect
to $g_E$.  In eqs. \eope\ we have taken the angular
average, which eliminates these singularities.)
This kind of complication is absent in the $(\phi^4)_4$
theory.
Hence, we can apply the variational formulas \evar\ to
evaluate
multiple derivatives of the correlation functions of composite

operators at $\lambda =0$.  In this way we can relate
the short distance singularities in the renormalized
theory to the singularities we encounter
in perturbative calculations.  Then, we can
interpret eqs.~\econsisttwo\ (or \eAfinal)
as a relation between the perturbative singularities
and anomalous dimensions.
Calculations of OPE coefficients for the
conjugate operators have been done
up to first order in $\lambda$.
\ref\rsonodaphi{H. Sonoda, ``Operator coefficients
for composite operators in the $(\phi^4)_4$ theory,''
UCLA preprint, UCLA/92/TEP/15}

\vskip .25in
I would like to thank Orlando Alvarez for discussions.

\vfill
\eject
\appendix{A}{Calculation of $\OE^* \Om^* - \Om^* \OE^*$}

In this appendix we will sketch the derivation of the
commutator
\ecommfinal.  From the definition \eEm\ we find, using \evar,
\eqn\eEmtwo{\eqalign{& (\partial_m \partial_{g_E} -
\partial_{g_E} \partial_m ) \vev{\O_i (0)}  \cr
& ~~ =
\lim_{\ep \to 0} \Bigg[ \int_{r \ge \ep} d^4 r ~
\lim_{\eta \to 0} \Big(
\int_{\scriptstyle |r' - r| \ge \eta \atop
\scriptstyle r'  \ge \eta} d^4 r' ~
\vev{(\Om (r') \OE (r) - \OE (r') \Om (r)) \O_i (0)} \cr
&~~~ + A_{m i,j} (\eta ) \vev{\OE (r) \O_j (0)}
- A_{E i,j} (\eta ) \vev{\Om (r) \O_j (0)} \Big) \cr
&~~~ - \int_{1 \ge r \ge \ep} d^4 r ~{\rm Sg}
\lim_{\eta \to 0}
\Big( {\rm ditto} \Big)
\Bigg] ~~
- \partial_m c_{Ei,j} \vev{\O_j} + \partial_{g_E} c_{mi,j}
\vev{\O_j} \cr
&~ + c_{Ei,j} \lim_{\ep \to 0}
\Big( \int_{r' \ge \ep} d^4 r'~ \vev{\Om (r') \O_j (0)}
+ A_{mj,k} (\ep ) \vev{\O_k} \Big) \cr
&~ - c_{mi,j} \lim_{\ep \to 0}
\Big( \int_{r' \ge \ep} d^4 r'~ \vev{\OE (r') \O_j (0)}
+ A_{Ej,k} (\ep ) \vev{\O_k} \Big) .\cr}}
Here ${\rm Sg}$ denotes the singular part with respect to the
coordinate
$r$.

Due to the antisymmetry of the integrand under interchange
of $m$ and $g_E$, the first integral over $r'$ can be
restricted to
$r' \le \ep$.  In the second integral over $r'$, we must keep
the entire range of integration, since the operation of taking
the
singular part spoils this antisymmetry.  Hence, we obtain
\eqn\einter{\eqalign{
&(\partial_m \partial_{g_E} - \partial_{g_E} \partial_m)
\vev{\O_i} \cr
& = \lim_{\ep \to 0}
\Bigg[ \int_{r \ge \ep} d^4 r~
\Bigg( \lim_{\eta \to 0}
\Big( \int_{\ep \ge r' \ge \eta} d^4r'~
\langle \OE (r) (\Om (r') \O_i (0) - C_{mi,j} (r') \O_j (0)) \cr
& ~~~~~~~~~~~~~~~~~~~~
- \Om (r) (\OE (r') \O_i (0) - C_{Ei,j} (r') \O_j (0) )
\rangle_{m,g_E}
\Big) \cr
& ~~~~~ + \int_{1 \ge r' \ge \ep} d^4 r'~
\Big( - C_{mi,j} (r') \vev{\OE (r) \O_j (0)} ~
+  C_{Ei,j} (r') \vev{\Om (r) \O_j (0)} \Big)
\Bigg)  \cr
& - \int_{r \ge \ep} d^4 r~ {\rm Sg} \lim_{\eta \to 0}
\Bigg( \int_{r' \ge \eta} d^4r'~
\vev{ (\OE (r) \Om (r')  - \Om (r) \OE (r')) \O_i (0) } \cr
& ~~~~~ + \int_{1 \ge r' \ge \eta} d^4 r'~
\Big( - C_{mi,j} (r') \vev{\OE (r) \O_j (0)} ~
+ C_{Ei,j} (r') \vev{\Om (r) \O_j (0)} \Big)
\Bigg)
\Bigg] \cr
& ~ + (\partial_{g_E} c_m - \partial_m c_E + [c_E, c_m])_{i,j}
\vev{\O_j} .\cr}}

Now using operator product expansions we find
\eqn\eopeone{\lim_{\ep \to 0} \int_{r \ge 1} d^4 r ~
\int_{\ep \ge r'} d^4 r'~
\vev{\OE (r) (\Om (r') \O_i (0) - C_{mi,j} (r') \O_j (0))} = 0}
and

\eqn\eopetwo{{\rm Sg} \int_{r' \ge 1} d^4 r'~
\vev{\Om (r') \OE (r) \O_i (0)}
= \int_{r' \ge 1} d^4 r'~ \vev{\Om (r') C_{Ei,j} (r) \O_j (0)} }
and similar equations obtained by interchanging $m$ and
$g_E$.

Substituting the above into \einter, we obtain
\eqn\eapfinal{\eqalign{&(\partial_m \partial_{g_E}
- \partial_{g_E} \partial_m) \vev{\O_i} \cr
& = \lim_{\ep \to 0} \Bigg[
\int_{1 \ge r \ge \ep} d^4 r~ \Big( \int_{1 \ge r' } d^4 r'~
\langle \OE (r) (\Om (r') \O_i (0) - C_{mi,j} (r') \O_j (0)) \cr
& ~~~~~~~~~~~~~~~ - \Om (r) (\OE (r') \O_i (0) - C_{Ei,j} (r')
\O_j (0) )
\rangle_{m,g_E} \Big) \cr
&~~~~~~~~~~ - \int_{1 \ge r \ge \ep} d^4 r ~ {\rm Sg}
\Big({\rm ditto}\Big) \Bigg] \cr
& ~~~+ (\partial_{g_E} c_m - \partial_m c_E + [c_E,
c_m])_{i,j} \vev{\O_j} .\cr}}
This gives \ecommfinal.

In the above result the range of the double integral is
restricted
to $r \le 1$, since we used $r=1$ as the renormalization point
when we introduced divergent subtractions $A_E, A_m$ in
\eA.  If we use
a different renormalization point $r = r_0 < 1$ to define the
subtractions $A_E, A_m$, then we must change the finite
counterterms $c_E, c_m$ to
\eqn\enewc{\eqalign{c'_E &= c_E - \int_{1 \ge r \ge r_0} d^4
r~
C_E (r) , \cr
c'_m &= c_m - \int_{1 \ge r \ge r_0} d^4 r~ C_m (r) .\cr}}
Then eq. \eapfinal\ can be rewritten as
\eqn\eapfinaltwo{\eqalign{&(\partial_m \partial_{g_E}
- \partial_{g_E} \partial_m) \vev{\O_i} \cr
& = \int_{r_0 \ge r} d^4 r~ {\rm F.P.} \int_{r_0 \ge r' } d^4 r'~
\langle \OE (r) (\Om (r') \O_i (0) - C_{mi,j} (r') \O_j (0)) \cr
& ~~~~~~~~~~ - \Om (r) (\OE (r') \O_i (0) - C_{Ei,j} (r') \O_j
(0)
\rangle_{m,g_E} \cr
& ~~~+ (\partial_{g_E} c'_m - \partial_m c'_E + [c'_E,
c'_m])_{i,j} \vev{\O_j} ,\cr}}
where ${\rm F.P.}$ denotes a finite part with respect to $r$.
We can take the radius $r_0$ as small as we want.  This
guarantees
the locality of the double integral.

In the presence of other composite operators the
consistency condition
\eapfinal\ is modified to
\eqn\eapfinalthree{\eqalign{&(\partial_m \partial_{g_E}
- \partial_{g_E} \partial_m) \vev{\On} \cr
& = \sum_{k=1}^n
\Bigg[ \int_{r_0 \ge |r-r_k|} d^4 r~
{\rm F.P.} \int_{r_0 \ge |r' - r_k| } d^4 r' \cr
& ~~~~~~~~~~~~~\langle \OE (r)
( \Om (r') \O_{i_k} (r_k) - C_{mi_k,j} (r'-r_k) \O_j (r_k)) \cr
& ~~~~~~~~~~~~~~~ - \Om (r)
( \OE (r') \O_{i_k} (r_k) - C_{Ei_k,j} (r'-r_k) \O_j (r_k))
\rangle_{m,g_E} \cr
& ~~~+ ( \partial_{g_E} c'_m - \partial_m c'_E + [c'_E, c'_m]
)_{i_k,j}
\vev{\O_{i_1} (r_1) ... \O_j (r_k) ... \O_{i_n} (r_n)}
\Bigg] ,\cr}}
where we can take $r_0$ smaller than the smallest distance
between
any two of the $n$ points $r_1, ..., r_n$.

\appendix{B}{Explicit calculation of $N(m,g_E)$}

We have found that the counterterms $c_m, c_E$ must
satisfy
the consistency condition \ecurv\
\eqn\ewhole{\partial_{g_E} c_m - \partial_m c_E + [c_E, c_m]
= \Omega_{Em} ,}
where the curvature matrix $\Omega_{Em}$ is defined by
eq. \eOmega.  By taking the maximal part we obtain

\eqn\emax{\partial_{g_E} \bar{c}_m - \partial_m \bar{c}_E
+ [\bar{c}_E, \bar{c}_m]
= \bar{\Omega}_{Em} ,}
where the bars denote the maximal parts.

Under a redefinition of the operators
\eqn\eredef{\O_i' = N_{i,j} (m,g_E) \O_j,}
where
\eqn\estructure{N_{i,j} (m,g_E) \propto m^{x_i - x_j} ,}
the finite counterterms transform as
\eqn\etrans{\eqalign{c'_m &= N c_m N^{-1} - \partial_m N
\cdot N^{-1} \cr
c'_E &= N c_E N^{-1} - \partial_{g_E} N \cdot N^{-1} . \cr}}
The non-maximal terms transform homogeneously, while the
maximal terms transform inhomogeneously:
\eqn\einhomo{\eqalign{\bar{c'}_m &= N \bar{c}_m N^{-1}
- \partial_m N \cdot N^{-1} \cr
\bar{c'}_E &= N \bar{c}_E N^{-1} - \partial_{g_E} N \cdot
N^{-1} . \cr}}

We wish to choose the matrix $N(m,g_E)$ such that
\eqn\egaugeone{\bar{c'}_E (m,g_E) = 0 }
\eqn\egaugetwo{\bar{c'}_m (m, 0) = 0 .}

The gauge condition \egaugeone\ is solved by
\eqn\efirst{N(m,g_E) = M(m)
\cdot {\cal R} \exp \left[ \int_0^{g_E} dx~\bar{c}_E (m,x)
\right] ,}
where ${\cal R}$ implies the increasing ordering of $x$
toward right,
and $M(m)$ is an arbitrary invertible matrix dependent only
on $m$.
The exponential is analytic in both $m$ and $g_E$.

Since
\eqn\eNM{N(m,0) = M(m) ,}
we find
\eqn\esecond{\bar{c'}_m (m,0) = M c_m (m,0) M^{-1} -
\partial_m M
\cdot M^{-1} .}
Hence, to satisfy the other gauge condition \egaugetwo, we
must
choose
\eqn\ethird{M(m) = L \cdot {\cal R} \exp \left[ \int_0^m dy
{}~c_m (y,0) \right] ,}
where $L$ is an arbitrary invertible constant matrix.  Note
that
$M(m)$ is analytic in $m$.

Finally we note that under the gauge conditions \egaugeone,
\egaugetwo,
we get a relation
\eqn\ecmomega{c'_m (m,g_E) = \int_0^{g_E}
dx~\Omega'_{Em} (m,x) .}

\listrefs
\bye